# A New Remote User Authentication Scheme Using Smart Cards with Check Digits

Amit K. Awasthi and Sunder Lal

**Abstract** — *In 2000, Hwang and Li proposed a new remote user authentication scheme using smart cards. Chan and Chang showed that the masquerade attack is successful on this scheme. Recently Shen, Lin and Hwang pointed out a different type of attack on this scheme and presented a modified scheme to remove these defects. . Further in 2003, Leung et al showed that this modified scheme is still vulnerable to the attack proposed by Chan and Cheng. In addition they showed that the extended attack proposed by Chang and Hwang also works well. Recently Kumar have suggested the idea of check digits to overcome the above attacks. In this paper we present a new scheme by using check digits which also overcomes these attacks and is more efficient than Kumar's scheme.* [1]

**Index Terms** — **Authentication, cryptography, security, cryptanalysis, smart cards.**

## I. INTRODUCTION

IN 1993, Chang and Wu [5] introduced Remote password authentication scheme with smart cards. A number of remote password authentication schemes with smart cards have been proposed since then. These schemes allow a valid user to login a remote server and access the services provided by the remote server.

In 2000, Hwang and Li [7] proposed a new remote user authentication scheme using smart cards, which was based on ElGamal's cryptosystem. Chan and Cheng [2] made a cryptanalysis of Hwang and Li's Scheme and pointed out that a legitimate user can easily create a valid pair of ($ID_j$, $PW_j$) without knowing the secrete key of the system '$x_s$'.

Later, Shen, Lin and Hwang [10] showed that a legitimate user can masquerade as another user. The user can compute some other user's password. They also proposed a modified scheme to patch up both 'Chen and Chang's attack' and their own attack. In the same year, Chang and Hwang [4] proposed an extended attack to solve the problem of Chan and Cheng's attack on Hwang and Li's authentication scheme.

[1] This work was supported in part by the University Grant Commission – INDIA under Grant No. 8-9(98)-(SR-I).
A. K. Awasthi is with the Department of Applied Science, Hindustan College of Science and Technology, Farah, Mathura, UP, INDIA. (E-mail: awasthi_hcst@ yahoo.com)
S. Lal is with the Department of Mathematics, Institute of Basic Science, Dr. B. R. A. University, Agra, UP, INDIA.

Communicated with COMNET, Elsevier Science

.

Further Leung et al [9] showed that weakness still exists in Shen, Lin and Hwang's scheme. By using similar attack of Cheng' attack or Chang and Hwang's attack a legitimate user can still impersonate other legal users to login a remote server.

Recently Kumar [8] suggested the idea of check digits to repair the Shen et al's scheme, which successfully removes the threats devised by Leung et al. In the same year Awasthi and Lal [1] also proposed a scheme which also removes these threats successfully.

In this paper we propose a new remote user authentication scheme, which is more efficient than Kumar's scheme in terms of computational cost.

## II. REVIEW OF THE HWNAG-LI SCHEME

In this section, we briefly review Hwang-Li scheme [5]. This scheme is composed of the initial phase, registration phase, the login phase and the authentication phase. Whenever a user registers itself through registration phase, the server identifies the user and issues a password and a smartcard, holding related information, through a secure channel. To access the remote server, user inserts its smart card in to the device and keys the password. The server authenticates the user in authentication phase. Different phases work as follows:

*A. Initial Phase*

The System Administrator (**SA**) generates the following parameters
  $p$ : a large prime number
  $h(.)$ : a one-way function
  $x_s$ : a secret key of the system

*B. Registration Phase*

A user $U_i$ who wants to register itself for accessing server services, submits it's $ID_i$ to the SA. SA computes $PW_i$ as
  $PW_i = ID_i^{x_s} \mod p$,

Registration center issues a password $PW_i$ and a smart card, which contains the public parameters ($h(.)$, $p$).

*C. Login Phase*

User $U_i$ inserts its smart-card and provides $ID_i$ and $PW_i$. The smart card then does the following:
1. Generate a random number $r$.
2. Compute $C_1 = ID_i^r \mod p$.
3. Compute $t = h(T \oplus PW_i) \mod p - 1$, where $T$ is the current time stamp.
4. Compute $M = ID_i^t \mod p$.
5. Compute $C_2 = M(PW_i)^r \mod p$.

6. Sends a message $C = (ID_i, C_1, C_2, T)$ to the remote system.

### D. Authentication Phase

The system receives the message $C$ at time $T'$, where $T'$ is the current time of system. Then the system does the following steps:
1. Test the validity of $ID_i$.
2. Test the validity of time interval between $T$ and $T'$.
3. Check whether the following equation holds:
$$C_2(C_1^{x_s})^{-1} = (ID_i)^{h(T \oplus PWi)} \mod p.$$
It is difficult for user $U_i$ to compute the secret key $x_s$ of the system from the equation:
$$PWi = ID_i^{x_s} \mod p.$$

## III. ATTACKS ON HWNAG-LI SCHEME

### A. Chan-Cheng's attack

According to this attack a legitimate user can easily create a valid pair of $(ID_j, PW_j)$ without knowing the secrete key of the system '$x_s$'.

Suppose Bob is a legal user with a pair of $(ID_b, PW_b)$ and he wants to create a valid pair of $(ID_j, PW_j)$ such that it satisfies
$$PW_j = ID_j^{x_s} \mod p.$$
He first computes $ID_j$ by
$$ID_j = (ID_b * ID_b) \mod p.$$
Then, he can computes $PW_j$ as follow:
$$PW_j = ID_j^{x_s} \mod p = (PW_b * PW_b) \mod p$$
As a result, Bob can create a valid pair of $(ID_f, PW_f)$ without knowing the secrete key $x_s$.

### B. Shen-Lin-Hwang's attack

According to this attack a legitimate user who wants to masquerade as another user $U_k$ may login a remote server and gain access.

Suppose Bob wants to masquerade as another user $U_k$. He first computes $ID_b$ by
$$ID_b = ID_k^r \mod p,$$
where $r$ is a random number such that $gcd(r, p) = 1$.
Then, he submits his $ID_b$ to register with the remote server. **SA** will verify identity and provide a smart card and password
$$PW_b = ID_b^{x_s} \mod p.$$
Now Bob can compute $PW_k$ as follow:
$$PW_k = ID_k^{x_s} \mod p = PW_b^{-r} \mod p$$
As a result, Bob can successfully use user $U_k$'s password to masquerade as the user $U_k$ to login the remote server.

### C. Chang and Hwang's attack

Chang and Hwang pointed out that the Chan and Cheng's attack is not always success. Chan and Cheng did not consider the validity checking of $ID_i$ in the authentication phase of Hwang and Li's scheme and cannot guarantee that the specific format of the user identity always satisfies the square of a legitimate identity, that is, $ID_i \mod p$ [3]. Chan and Hwang also designed two attack mechanisms to obtain a valid identity which is based on Chan and Chang's one.

*I-Mechanism*
Following Chan and Chang' equation
$$ID_j = (ID_b * ID_b) \mod p.$$
is modified to:
$$ID_j = (ID_b)^r \mod p.$$
where $r$ is an arbitrary integer and the corresponding password can be obtained by:
$$PW_j = (PW_b)^r \mod p.$$
If $ID_b$ is primitive element of the modulus $p$, any valid identity $ID_j$ as well as its password can be computed. And now any intruder with identity $ID_j$ successfully passes the validity check for identity in Hwang and Li's scheme.

*II-Mechanism*
Several intruders may cooperate to get a valid identity $ID_j$ as well as its corresponding password $PW_j$ as follows
$$ID_j = \prod ID_b \mod p.$$
Suppose identity fulfills the required format, its corresponding password can be obtained by:
$$PW_j = (ID_j)^{x_s} \mod p.$$
$$= \prod PW_b \mod p.$$

## IV. SHEN-LIN-HWANG'S MODIFICATION

Shen, Lin and Hwang proposed a modified scheme in [7]. They suggested that instead of an identity a shadow identity will be given to the legal user. The scheme has the just hiding of identities to prevent from the forgery attack [section III-B, III-C]. The steps in login and authentication phase are not changed except $ID_i$ is replaced by $SID_i$. The modified registration phase is as follows:

User $U_i$ submits his $ID_i$ to the remote server for registration. The remote server computes the pair $(SID_i, PW_i)$ as follows:
$$SID_i = Red(ID_i)$$
$$PW_i = (SID_i)^{x_s} \mod p.$$
Where $Red(.)$ is a shadow function, such that user cannot compute $ID_i$ having the knowledge of $SID_i$ and $Red(.)$. The shadow function is only maintained in the remote server. $SID_i$ is user's shadow identity which is made public instead of $ID_i$. Registration center issues a shadow identity $SID_i$, a password $PW_i$ and a smart card, which contains the public parameters ($h(.), p$). In login phase and authentication phase each $ID_i$ is replaced with $SID_i$. Login and authentication processes remain same. Now message C coming to server through login page is $(SID_i, C_1, C_2, T)$ instead of $(ID_i, C_1, C_2, T)$.

According to Shen et al., Though shadow identity of $ID_i$ is known, shadow identity of $ID_b = ID_k^r \mod p$ (or $ID_b = ID_k * ID_k \mod p$) can not be computed. That is why this scheme is secure.

## V. LEUNG-CHENG-FONG-CHAN'S OBJECTION

Leung, Cheng, Fong and Chan pointed out that although Shen et al. defend the attack of registration of a new $ID_j$ via $ID_i$ from a legal user $U_i$, the scheme can not withstand attack that similar to Chan and Cheng's attack and Chang and Hwang's attack in authentication phase.

According to Leung et al, shadow identity does not enhance the security in authentication phase and hence intruder can still login to the remote system by having a pair ($SID_j$, $PW_j$). Attack works as follows:

Replace $ID_j$ with $SID_j$ in Chang and Hwang's attack (mechanism-I), we get
$$SID_j = (SID_b)^r \mod p.$$
where $r$ is an arbitrary integer and the corresponding password can be obtained by:
$$PW_j = (SID_j)^{x_s} \mod p.$$
$$= [(SID_b)^r]^{x_s} \mod p.$$
$$= (PW_b)^r \mod p.$$

This shows that a legitimate user ui can impersonate other legal user with valid pair ($SID_j$, $PW_j$) to login the remote server. Note $SID_b$ is primitive element of the modulus p, any valid shadow identity $SID_j$ as well as its password can be computed. Similarly, chan-cheng's attack works fine.

## VI. REVIEW OF THE KUMAR'S SCHEME

In this section, we briefly review Kumar's scheme [8]. This scheme is composed of the initial phase, registration phase, the login phase and the authentication phase.

This scheme is the modified form of the Shen- Lin- Hwang's scheme and uses one more function $C_K$ to generate the check digit [6] for each registered identity. In this scheme, only the AS can generate a valid identity and the corresponding *check digit*.

### A. Initialization Phase

In this phase, the *AS* generates the following parameters
$P$ : a large prime number.
$f$ : a one-way function.
$x_s$: a secret key of the system, which is only possessed with the *AS*.
*Red* (.): a "*shadowed*" identity of the device which is only possessed with the *AS*.
$C_K$ (.): a one way secure check digit function, which is used to generate unique *check digit* for the registered identity, and also is possessed with the *AS* Only.

### B. Registration Phase

Assume that this phase is executed over a secure channel. User $U$ submits her/his identity string $ID_i$ to the *AS*. The string $ID_i$ consists of the name of the user and a unique identification number etc, which are unique for the user $U$. Then, *AS* computes the following:

$S_{ID} = Red(ID_i)$, $C_{ID} = C_K(S_{ID})$ and $PW = (S_{ID})^{x_s} \mod p$.

Furthermore, the *AS* distributes the smart card and ($S_{ID} \| C_{ID}$, $PW$) to the user $U$ in a secure way (say physically). The smart card contains the public parameters ($f$, $p$).

### C. Login Phase

The user $U$ attaches her/his smart card to the smart card reader at any time $T$ and keys her/his identity $S_{ID} \| C_{ID}$ and password $PW$. The smart card conducts the following computations:

1. Generate a random number $r$.
2. Computes $C_1 = (S_{ID})^r \mod p$.
3. Compute $t = f(T \oplus PW) \mod p - 1$.
4. Compute $m = (S_{ID})^t \mod p$.
5. Compute $C_2 = m(PW)^r \mod p$.
6. Sends a message $L_R = (S_{ID} \| C_{ID}, C_1, C_2, T)$ to the *AS*.

### D. Authentication Phase

Assume that the *AS* receives the login request $L_R$ at time $T_c$. Then, *AS* does the following computations to check the validity of the login request $L_R$.

1. Check the specific format of $S_{ID}$. If the format of the $S_{ID}$ is incorrect, then *AS* rejects the login request $L_R$.
2. Check, whether the condition $C_{ID} = C_K(S_{ID})$ holds, if not, then *AS* rejects the login request $L_R$.
3. Check, whether $T_c - T \leq \Delta T$, where $\Delta T$ is the legal time interval due to transmission delay, if not, then *AS* rejects the login request $L_R$.
4. Check, if $C_2 \stackrel{?}{=} (C_1^{x_s})(S_{ID})^{f(T \oplus PW)} \mod p$, then the *AS* accepts the login request. Otherwise, the login request will be rejected by *AS*.

## VII. A NEW REMOTE USER AUTHENTICATION SCHEME USING SMART CARDS WITH CHECK DIGITS

In this section, we propose our new remote user authentication scheme. This scheme is also composed of the initial phase, registration phase, the login phase and the authentication phase.

This scheme is the modified form of the Hwnag-Li's scheme and uses one more function $C_K$ to generate the check digit [8] for each registered identity. In this scheme, only the *AS* can generate a valid identity and the corresponding *check digit*.

### A. Initialization Phase

The System Administrator (**SA**) generates the following parameters
$p$ : a large prime number
$f(.)$ : A one-way function
$x_s$ : a secret key of the system
$C_K(.)$: a "function" to generate *check digit* for the registered identity, which is only possessed with the *AS*.

### B. Registration Phase

A user $U_i$ who wants to register itself for accessing server services, submits it's $ID_i$ to the SA. SA assigns a unique registration number $R$ and computes $PW_i$ as
$$PW_i = (ID_i \oplus R)^{x_s} \mod p,$$
$$C_{ID} = C_K(ID_i \oplus R)$$

Registration center issues a password $PW_i$ and a smart card, which contains the public parameters ($h(.)$, $p$, $R$, $C_{ID}$).

### C. Login Phase

The user $U$ attaches its smart card to the smart card reader at any time $T$ and keys the identity $ID_i$ and password $PW_i$. The smart card computes as foloowing:

1. Generate a random number $r$.

2. Computes $C_1 = (ID_i \oplus R)^r \mod p$.
3. Compute $t = f(T \oplus PW_i) \mod p - 1$.
4. Compute $m = (ID_i)^t \mod p$.
5. Compute $C_2 = m (PW_i)^r \mod p$.
6. Sends a message $C = (ID_i, R, C_{ID}, C_1, C_2, T)$ to the AS.

### D. Authentication Phase

The system receives the message $C$ at time $T'$, where $T'$ is the current time of system. Then the system does the following steps:

1. Test the validity of $ID_i$.
2. Checks $C_{ID} = C_K(ID_i \oplus R)$, if true then proceed else reject the request.
3. Test the validity of time interval between $T$ and $T'$, in case of invalid time interval give an appropriate message and stop the protocol.
4. Check whether the following equation holds:
$$C_2(C_1^{x_S})^{-1} = (ID_i)^{h(T \oplus PW_i)} \mod p.$$

## VIII. SECURITY ANALYSIS OF THE SCHEME

In this section, we shall only discuss the enhanced security features. Rests are the same as original schemes in literature.

### A. Chan- Cheng's attack and Chang- Hwang's Attack

The security explanation is very similar to Kumar's discussion. The functionality of the $C_K(.)$ blocks the attacks via authentication phase as in *Chan- Cheng's attack and Chang- Hwang's Attack*. Assume, Alice chooses a $R'$ such that $ID_b \oplus R' = (ID_i \oplus R)^k \mod p$, where $k$ is a random number. Then, he can compute the corresponding password
$$PW_b = PW_i^k \mod p.$$

But, this result is incomplete; still, it is essential to obtain the check digit corresponding to $ID_b$. In the proposed scheme only the SA can generate the check digit corresponding to $ID_b$ and $R'$. As a result, a legal user $ID_i$ cannot compute a valid pair of identity and password for the other user $ID_b$ to login at the server. Thus, *Chang- Hwang's Attack* will not work. As being the one of the case of this attack, Chan–Cheng's attack also does not work.

### A. Shen-Lin-Hwang's attack

The functionality of the *assignment of registration number* blocks the masquerade attack via registration phase. Assume that an intruder intercepts the logon request $C = (ID_i, C_{ID}, C_1, C_2, T_c)$ from a public network and computes
$$ID_b = (ID_i)^k \mod p.$$
Now he submits this $ID_b$ to the SA for the registration. Upon receiving the registration request from $ID_b$, the SA will reject the request because the format of $ID_b$ is incorrect. If in any case format holds, SA computes assigns a unique registration number $r$ and computes a corresponding password $PW_b = (ID_i \oplus r)^{x_s} \mod p$. With this $PW_b$ it is infeasible to compute $PW_i$, because R (Registration number of user $U_i$) and r (Registration number of user $U_b$ corresponding to $ID_b$) are predefined.

### B. Leung-Cheng-Fong-Chan's attack:
is extension of *Chan-Cheng's attack and Chan and Hwang's attack*. In this scheme we are not using shadow function so there is no need to discuss this attack.

## IX. COMPARISON

In this section we compare the computational cost of our scheme with other schemes. Let us denote E as computational cost of an exponential operation, H as cost of hashing, M cost of multiplication, R is cost of redirection function (It is approximately hashing cost) and C as Check digit counting cost. Since R and C both functions are equivalent to hash function, we can not neglect their computational cost. Other costs are negligible or minor. Then following table shows the computational cost in various phases in terms of these costs.

Table 1. Comparison of some remote user authentication schemes .

| Schemes | Computation Cost | | | Security in authentication and registration phases |
|---|---|---|---|---|
| | Registration Phase | Login Phase | Authentication Phase | |
| Hwang-Li Scheme | E | 3E+H+M | 3E+H+M | NIL |
| Shen et al's scheme | R+E | 3E+H+M | 3E+H+M | NIL |
| Kumar's Scheme | R+E+C | 3E+H+M | 3E+H+M+C | Yes |
| Proposed Scheme | E+C | 3E+H+M | 3E+H+M+C | Yes |

Above table shows that our proposed scheme bears a less redirection cost in registration phase than Kumar's scheme and holds similar security level.

## X. COMPARISON

This paper presents *New Remote User Authentication Scheme Using Smart Cards with check digits*. This scheme is more efficient than Kumar's scheme. The proposed scheme is secure against the both types of attacks: attacks via registration phase and the attack via authentication phase. One enhanced feature of this scheme is that more than one user with same identity may register to the server.

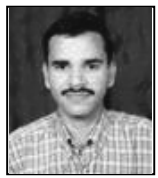
**Amit K Awasthi** received his M. Sc. Degree in 1999 from Bareilly College, (M. J. P. Rohilkhand University,) Bareilly. He is currently a lecturer in Department of Applied Science, Hindustan College of Science and Technology, Farah, Mathura, INDIA. He is member of Indian Mathematical Society, Group for Cryptographic Research, Cryptography Research Society of India and Computer Society of India. His current research interests include data security and cryptography.

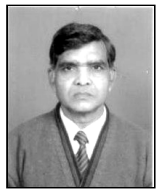
**Sunder Lal** is currently a professor and Head of Department of Mathematics, IBS Khandari, Dr. B. R. A. University, Agra, INDIA. He is member of Indian Mathematical Society, Group for Cryptographic Research, and Cryptography Research Society of India. His current research interests include cryptography, number theory and applied algebra.